\documentclass[final,3p,times]{elsarticle}

\usepackage{amssymb}
\usepackage{amsmath}
\usepackage{multirow}

\newcommand\mat\mathbf
\newcommand\dd{\, \mathrm{d}}
\newcommand\ii{\mathrm{i}}

\newcommand\um{~{\mu}\textrm{m}}
\newcommand\nm{\textrm{~nm}}
\newcommand\mm{\textrm{~mm}}

\makeatletter
\def\ps@pprintTitle{%
  \let\@oddhead\@empty
  \let\@evenhead\@empty
  \let\@oddfoot\@empty
  \let\@evenfoot\@oddfoot
}
\makeatother

\begin{document}

\begin{frontmatter}

\title{Designing robust diffractive neural networks with improved transverse shift tolerance}

\author[label2,label1]{Daniil~V.~Soshnikov}
\author[label1,label2]{Leonid~L.~Doskolovich}
\author[label2,label1]{Georgy~A.~Motz}
\author[label2,label1]{Egor~V.~Byzov}
\author[label2,label1]{Evgeni~A.~Bezus}
\author[label1,label2]{Dmitry~A.~Bykov}

\affiliation[label2]{organization={Samara National Research University},
            addressline={34 Moskovskoye shosse}, 
            city={Samara},
            postcode={443086}, 
            country={Russia}}
						
\affiliation[label1]{organization={Image Processing Systems Institute, National Research Centre ``Kurchatov Institute''},
            addressline={151 Molodogvardeyskaya st.}, 
            city={Samara},
            postcode={443001}, 
            country={Russia}}

\begin{abstract}
A wide range of practically important problems is nowadays efficiently solved using artificial neural networks.
This gave momentum to intensive development of their optical implementations, among which, the so-called diffractive neural networks (DNNs) constituted by a set of phase diffractive optical elements (DOEs) attract considerable research interest.
In the practical implementation of DNNs, one of the standing problems is the requirement for high positioning accuracy of the DOEs.
In this work, we address this problem and propose a method for the design of DNNs for image classification, which takes into account the positioning errors (transverse shifts) of the DNN elements.
In the method, the error of solving the classification problem is represented by a functional depending on the phase functions of the DOEs and on random vectors describing their transverse shifts.
The mathematical expectation of this functional is used as an error functional in the gradient method for calculating the DNN taking into account the transverse shifts of the DOEs.
It is shown that the calculation of the derivatives of this functional corresponds to the DNN training method, in which the DOEs have random transverse shifts.
Using the proposed gradient method, DNNs are designed that are robust to transverse shifts of the DOEs and enable solving the problem of classifying handwritten digits at a visible wavelength.
Numerical simulations demonstrate good performance of the designed DNNs at transverse shifts of up to 17 wavelengths.

\end{abstract}


\end{frontmatter}

\section{Introduction}
Artificial neural networks have attained great success in solving various machine learning problems, including image classification and analysis, medical diagnostics, speech recognition, solving various inverse problems arising in nanophotonics, microscopy, holography, etc.
This, in particular, gave momentum to intensive development of the optical implementation of neural networks~\cite{1,2,3,4,5}, which is promising for overcoming the limitations of electronic components, which approach their physical limits, and for obtaining high energy efficiency and high-performance computing at the speed of light.
Among different platforms, the so-called diffractive neural networks (DNNs) attract considerable research interest.
Such neural networks are constituted by a cascade of sequentially located phase diffractive optical elements (DOEs)~\cite{6,7,8,9,10,11,12,13,14,15,16,17,18,19,20,21,22}.
It should be noted that the scientific field dedicated to the design of DOEs (including cascaded ones) for solving various problems of steering and shaping laser radiation has a long history and has been actively developing for more than 50 years~\cite{23,24,25,26,27,28,29,30,31}.
At the same time, the use of cascaded DOEs for optical solution of machine learning problems was first demonstrated only in 2018 in the seminal work~\cite{6}.
In this work, using a number of analogies between a cascaded DOE and a conventional artificial neural network, the term “diffractive (deep) neural network” was proposed and the possibility of optically solving classification problems using cascaded DOEs was theoretically and experimentally demonstrated.
Later works examined the use of DNNs (or cascaded DOEs) for solving various classification problems~\cite{7, 8, 10,11,12,13, 16,17,18,19}, object and video recognition~\cite{12, 14}, identification of the main object in an image (referred to as the salient object)~\cite{8}, implementation of various mathematical transformations described by linear operators~\cite{10, 17, 20} (including the case of parallel operation with several working wavelengths~\cite{21}), and imaging~\cite{22}.
The main method for the DNN design is the stochastic gradient descent method, as well as its various modifications~\cite{32}.

In the practical implementation of DNNs, one of the standing problems is the requirement for high positioning accuracy of the DOEs constituting the DNN.
In this regard, in many works, experimental studies of DNNs are carried out on a terahertz (THz) platform~\cite{6, 9, 11, 21}.
In particular, in the basic work~\cite{6}, experimental investigation of a DNN comprising five DOEs and solving the problem of classifying handwritten digits was carried out using a THz source with a wavelength of 0.75~mm.
At this wavelength, it was relatively easy to position the DOEs with small errors as compared to both the operating wavelength and the pixel size of the diffractive microrelief, which was equal to $0.4\mm$.
Note that the DOEs in~\cite{6} were placed in a special holder providing a positioning accuracy of $0.1\mm$.
For the case of visible radiation, the need for precise positioning of DOEs becomes a much more complex problem, especially with a large number of DOEs~\cite{10, 12, 13}.
In particular, in~\cite{13}, a DNN consisting of five DOEs (the pixel size of the diffractive microrelief being $4\um$) was theoretically and experimentally studied for the optical classification of handwritten digits at $632.8\nm$.
In that work, it was shown that random DOE positioning errors, namely, transverse shifts by a single pixel lead to a strong decrease in the resulting classification accuracy (from 91.2\% to 67.6\%).
At the same time, the longitudinal shifts of the DOE have a much smaller effect on the DNN performance~\cite{13}.
Note that to position the DOE in~\cite{13}, high-precision motorized platforms with a step of $1\um$ along the coordinate axes were used.
As a result, a classification accuracy of 88\% was experimentally achieved (vs.\ 91.2\% obtained in the simulations).

One of the ways to solve the problem of the positioning accuracy, which can be used in the case when phase spatial light modulators (SLMs) are utilized as DOEs constituting the DNN, is to train the DNN on fields measured in the implemented optical setup~\cite{10, 12}.
However, to measure the amplitudes and phases of the light fields incident on the SLMs, a complex setup is required, including a set of interferometers (their quantity being equal to the number of the utilized SLMs).
In this case, it is also necessary to repeatedly solve the ill-posed problem of reconstructing the phase of the light field from the measured interferograms.
The present authors believe that the most promising approach to tackle the problem of the positioning accuracy of the DNN elements is the use of special design methods that take into account the DOE positioning errors at the design stage.
In particular, in~\cite{11}, a method for training a DNN was proposed, in which the positions of the DOEs at the training stage were specified with random errors.
It was shown that this approach enables avoiding a sharp decrease in the classification accuracy at relatively large transverse shifts of the DOEs amounting to about 8.5 wavelengths.
At the same time, this work did not provide a theoretical justification for such a training strategy.
In addition, the results presented in~\cite{11} were obtained for the THz frequency range.

In this work, we consider a gradient descent method for designing DNNs, which are engineered to be robust to random positioning errors, namely, the transverse shifts of the DOEs constituting the DNN.
In what follows, we will refer to such DNNs as \emph{robust DNNs}.
In the method, the error of solving the classification problem is represented as a functional, which depends not only on the phase functions of the DOE, but also on random vectors describing the DOE shifts.
The mathematical expectation of this functional depending only on the DOE phase functions is then used as the error functional in the DNN design problem.
We obtain explicit expressions for the derivatives of this error functional taking into account the transverse shifts.
It is shown that the calculation of these derivatives using the Monte Carlo method corresponds to the DNN training method, in which the DOEs are positioned with random transverse shifts.
Using the proposed gradient method, we design DNNs consisting of a single DOE and a cascade of two DOEs operating at the 632.8~nm wavelength and having the pixel size (discretization step of the DOE phase) of $4\um$.
The designed DNNs perform optical classification of handwritten digits from the MNIST database.
The numerical simulation results demonstrate that the designed robust DNNs retain their good performance if the DOEs are shifted by 1--2 pixels along both coordinate axes (this shift distance corresponds to 6--17 operating wavelengths).
At the same time, the non-robust DNN (DNN designed \emph{without} taking into account the DOE shifts) comprising two DOEs becomes inoperable already when the DOEs are shifted by a single pixel (by $4\um$).

The paper is organized as follows.
In Section~\ref{sec:1}, we consider the problem statement of designing a DNN for optical image classification and calculate the derivatives of a general error functional, which does not take into account the transverse shifts of the DOEs constituting the DNN.
In Section~\ref{sec:2}, using the results of Section~\ref{sec:1}, we formulate a method for the robust DNN design taking into account the transverse shifts.
In Section~\ref{sec:3}, we design robust and non-robust DNN examples and compare their performance.

\section{Design of non-robust DNNs without taking into account the DOE shifts}\label{sec:1}
\begin{figure}[hbt]
	\centering
		\includegraphics{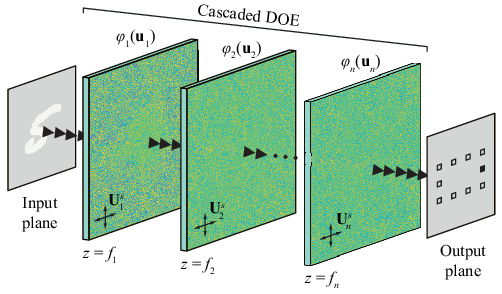}
	\caption{\label{fig:1} Geometry of the DNN design problem for solving a classification problem. The arrows schematically show random transverse shifts of the DOEs constituting the DNN.}
\end{figure}

Let us consider the problem of calculating a DNN (cascaded DOE) intended for the solution of a certain image classification problem.
Let the DNN consist of $n$ phase DOEs located in the planes $z = f_1, \ldots, z = f_n$ $(0 < f_1 < \cdots < f_n)$ and defined by the phase functions $\varphi_1( \mat{u}_1 ), \ldots, \varphi_n( \mat{u}_n )$, where $\mat{u}_j = ( u_j, v_j )$ are Cartesian coordinates in the planes $z = f_1, \ldots, z = f_n$ (Fig.~\ref{fig:1}). 
We assume that in the input plane $z = 0$, amplitude images of objects belonging to $N$~different classes are sequentially generated.
Each generated image is illuminated by a plane wave with wavelength $\lambda$. 
Let us denote by $w_{0,j}(\mat{u}_0)$ the complex amplitude of the light field generated in the input plane and corresponding to an object of the $j$-th class. 
Here and in what follows, the subscripts of the complex amplitude denote the index of the plane, in which this amplitude is written, and the class of the input image.

The light field $w_{0,j}(\mat{u}_0)$ generated at $z = 0$ then propagates through the cascaded DOE to the output plane $z = f_{n+1}$.
We assume that the propagation of the field in the free space (between the planes, in which the DOEs are located) is described by the Fresnel–Kirchhoff integral of the scalar diffraction theory and that the transmission of the field through a DOE is described in the thin optical element approximation as multiplication of the complex amplitude of the field by the complex transmission function (CTF) of the DOE, which, for the $m$-th DOE, has the form
\begin{equation}
\label{eq:1}
T_m(\mat{u}_m) = \exp \left\{ \ii{\varphi_m}(\mat{u}_m) \right\}.
\end{equation}
Under the made assumptions, the propagation of the input beam $w_{0,j}( \mat{u}_0)$ from the input plane $z = 0$ through the cascaded DOE to the output plane $z = f_{n+1}$ is described by the following recurrent formula:
\begin{equation}
\label{eq:2}
w_{m,j}( \mat{u}_m ) 
	= 
	C_m\iint w_{m - 1,j}( \mat{u}_{m-1} )T_{m - 1,q}( \mat{u}_{m-1} )
	\cdot \exp \left\{ \ii\frac{k_0}{2 d_m} ( \mat{u}_m - \mat{u}_{m-1} )^2 \right\}\dd^2\mat{u}_{m-1},
	m = 1,\ldots ,n + 1, 
\end{equation}
where $w_{m,j}(\mat{u}_m),\,m = 1,\ldots,n$ are the complex amplitudes of the field incident on the DOEs with the CTFs $T_m(\mat{u}_m)$, $k_0 = 2\pi/ \lambda$ is the wavenumber, $C_m = (\ii \lambda d_m)^{-1}\exp\{\ii k_0 d_m\}$, and $d_m = f_m - f_{m-1}$ are the distances between planes.
For calculating the field $w_{1,j} ( \mat{u}_1 )$ incident on the first DOE [$m = 1$ in Eq.~\eqref{eq:2}], $T_0(\mat{u}_0) \equiv 1$ has to be used in Eq.~\eqref{eq:2}.

Let us assume that in the output plane $z = f_{n + 1}$, $N$ spatially separated target regions $G_k, k = 1,\ldots,N$ are defined, which correspond to $N$ different classes of the classification problem being solved.
For each input image, certain ``energies'' $E_k,\,k = 1, \ldots, N$ are generated in these regions.
Under the energies, we will understand either the integrals of the generated intensity distribution $I_{n + 1,j}(\mat{u}_{n + 1}) = {\left| w_{n + 1,j}(\mat{u}_{n + 1}) \right|^2}$ over the target regions
\begin{equation}
\label{eq:3}
E_k = \iint I_{n + 1,j}(\mat{u}_{n + 1})\chi_k(\mat{u}_{n + 1}) \dd^2 \mat{u}_{n + 1},
\end{equation}
or the maximum intensity values in the target regions
\begin{equation}
\label{eq:4}
E_k = \mathop {\max}\limits_{\mat{u}_{n + 1} \in G_k} I_{n + 1,j}( \mat{u}_{n + 1} ) = I_{n + 1,j}( \mat{u}_{n + 1,k}),
\end{equation}
where $\chi_k(\mat{u}_{n + 1})$ is the indicator function of the region $G_k$ and $\mat{u}_{n + 1,k} = \mathop {\arg \max }\nolimits_{{\mat{u}_{n + 1}} \in G_k} I_{n + 1,j}( \mat{u}_{n + 1} )$ are the coordinates of the maxima.
In order to solve the classification problem, it is necessary for the cascaded DOE to generate such an energy distribution in the output plane for an input field $w_{0,j}(\mat{u}_0)$ (an object of the $j$-th class), so that the maximum of the generated energies ${E_k}$ [Eq.~\eqref{eq:3} or Eq.~\eqref{eq:4}] is reached in the $j$-th target region $G_j$~\cite{6,7,8}.
Thus, the design of a DNN can be regarded as the problem of finding the phase functions $\varphi_1( \mat{u}_1 ), \ldots, \varphi_n(\mat{u}_n )$ of the DOEs constituting the DNN, which provide the above described operation. 
Let us note that  if the DNN is implemented as diffractive microrelief, its
% for the practical implementation of the DNN, the functions of the diffractive microrelief 
height functions can be calculated from the phase functions as
\begin{equation}
\label{eq:5}
h_m( \mat{u}_m ) = \frac{\varphi_m(\mat{u}_m)}{k_0 (n_r - 1)},
\end{equation}
where $n_r$ is the refractive index of the DOE material.

Following~\cite{6,7,8, 10,11,12,13}, for the DNN design, we will use a stochastic gradient descent method in the neural network paradigm.
Let us start with a general description of the method.
We assume that for the calculation (training) of the cascaded DOE, a training set is used, which contains a set of input distributions (complex amplitudes of the fields) corresponding to the images of objects of different classes.
At each step of the method, a number of input distributions (referred to as batch) is randomly chosen.
For the chosen batch, the derivatives of the error functional $\varepsilon (\varphi_1, \ldots, \varphi_n )$, which depends on the phase functions and evaluates the DNN operation, are calculated and a step in the direction of the anti-gradient is performed.
Without loss of generality, we will assume that the batch contains certain input distributions $w_{0,j}(\mat{u}_0),\,\,j = 1,\ldots,N$ from all $N$ different classes.
To describe the calculations performed for each batch, let us write the error functional in an explicit form.
Let the classification error of the input beam $w_{0,j}(\mat{u}_0)$ representing an object from the $j$-th class be defined by a certain error functional $\varepsilon_j(\varphi_1, \ldots, \varphi_n )$.
Since the classification is performed by analyzing the distribution of the energies $E_k$, calculated using Eq.~\eqref{eq:3} or Eq.~\eqref{eq:4}, the general form of the functional $\varepsilon_j( \varphi_1,\ldots,\varphi_n )$ is
\begin{equation}
\label{eq:6}
\varepsilon_j( \varphi_1, \ldots, \varphi_n ) = D_j(E_1, \ldots, E_N),
\end{equation}
where $D_j(E_1, \ldots, E_N)$ is a certain function describing the deviation of the generated energy distribution from the required one, in which the maximum of the energy is reached in the “correct” $j$-th region.
Then, the error functional for a batch containing distributions $w_{0,j}(\mat{u}_0),\,\,j = 1,\ldots,N$ can be represented as a sum of the functionals introduced above:
\begin{equation}
\label{eq:7}
\varepsilon (\varphi_1,\ldots,\varphi_n) = \sum\limits_{j = 1}^N \varepsilon_j(\varphi_1,\ldots,\varphi_n).
\end{equation}

For the functional of Eq.~\eqref{eq:7}, one can easily calculate the Fréchet derivatives $\delta \varepsilon / \delta {\varphi_m}$.
Indeed, since the functional~\eqref{eq:7} equals the sum of functionals, its derivatives have the form
\begin{equation}
\label{eq:8}
\frac{{\delta \varepsilon ( \varphi_1,\ldots,\varphi_n )}}{{\delta {\varphi_m}}} = \sum\limits_{j = 1}^N {\frac{{\delta {\varepsilon_j}(\varphi_1,\ldots,\varphi_n)}}{{\delta {\varphi_m}}}} ,\,\,m = 1,\ldots,n.
\end{equation}

Let us calculate the derivative $\delta \varepsilon_j / \delta \varphi_m$ from Eq.~\eqref{eq:8}.
For this, let us write the increment of the functional $\varepsilon_j( \varphi_1,\ldots,\varphi_n)$ caused by an increment $\Delta \varphi_m$ of the phase function of the $m$-th DOE as
\begin{equation}
\label{eq:9}
\Delta_m\varepsilon_j( \varphi_1,\ldots,\varphi_n ) = \varepsilon_j( \varphi_1,\ldots,\varphi_m + \Delta \varphi_m,\ldots,\varphi_n ) - \varepsilon_j( \varphi_1,\ldots,\varphi_m,\ldots,\varphi_n ).
\end{equation}
Similarly to~\cite{18, 19}, this increment can be represented as the following scalar product:
\begin{equation}
\label{eq:10}
\begin{aligned}
  \Delta_m{\varepsilon_j}( \varphi_1,\ldots,\varphi_n ) 
	&= 
	2\operatorname{Re}  \iint {\left[ {{\Delta_m}{w_{n + 1,j}}({\mat{u}_{n + 1}})} \right]\,F_{n + 1,j}^*({\mat{u}_{n + 1}}){\dd^2}{\mat{u}_{n + 1}}} 
\\&= 
2\operatorname{Re} \left\langle {{\Delta_m}{w_{n + 1,j}}({\mat{u}_{n + 1}}),{F_{n + 1,j}}({\mat{u}_{n + 1}})} \right\rangle, 
\end{aligned}
\end{equation}
where $\Delta_m w_{n + 1,q,j} ( \mat{u}_{n + 1} )$ is the increment of the complex amplitude caused by the phase increment $\Delta \varphi_m$, the angle brackets denote the scalar product of functions, and the function $F_{n + 1,j}( \mat{u}_{n + 1} )$ depends on both the optimization criterion [the form of the functions $D_j(E_1,\ldots,E_N)$] and the definition of energies $E_k$.
If under the energies, the integrals of the intensity over the target regions are understood [see Eq.~\eqref{eq:3}], then the function $F_{n + 1,j}( \mat{u}_{n + 1} )$ has the form~\cite{18}
\begin{equation}
\label{eq:11}
{F_{n + 1,j}}\left( \mat{u}_{n + 1} \right) = w_{n + 1,j}(\mat{u}_{n + 1}) \sum\limits_{k = 1}^N {\chi_k(\mat{u}_{n + 1})\frac{\partial D_j}{\partial E_k}} .
\end{equation}
In the case, in which the maximum values of the intensity in the target regions are used as the energies [see Eq.~\eqref{eq:4}], the function $F_{n + 1,j}( \mat{u}_{n + 1} )$ is calculated as~\cite{19}
\begin{equation}
\label{eq:12}
F_{n + 1,j}( \mat{u}_{n + 1} ) = w_{n + 1,j}(\mat{u}_{n + 1}) \sum\limits_{k = 1}^N {\delta (\mat{u}_{n + 1} - \mat{u}_{n + 1,k})\frac{\partial D_j}{\partial E_k}} ,
\end{equation}
where $\delta ( \mat{u}_{n + 1} )$ is the delta function, and $\mat{u}_{n + 1,k} = \mathop {\arg \max }\nolimits_{{\mat{u}_{n + 1}} \in G_k} I_{n + 1}( \mat{u}_{n + 1} )$.
In what follows, we will refer to the function $F_{n + 1,j}( \mat{u}_{n + 1} )$ as the error field.

It is important to note that the operators describing the forward propagation [see Eq.~\eqref{eq:2}] and the backpropagation of the light field through a cascade of phase DOEs are unitary and conserve the scalar product~\cite{15}.
Using the scalar product conservation property, let us represent the increment~\eqref{eq:10} of the error functional as
\begin{equation}
\label{eq:13}
\Delta_m\varepsilon_j( \varphi_1,\ldots,\varphi_n ) = 2\operatorname{Re} \left\langle {{\Pr }_{f_{n + 1} \to f_m^+ }({\Delta_m}{w_{n + 1,j}}),{\Pr}_{f_{n + 1} \to f_m^+ }({F_{n + 1,j}})} \right\rangle ,
\end{equation}
where ${\Pr_{f_{n + 1} \to f_m^+ }}$ is the backpropagation operator of the field from the output plane $z = f_{n + 1}$ to the plane $z = f_m^+$ located immediately after the plane of the $m$-th DOE $z = f_m$.
Note that the backpropagation of the field, just as the forward one, is described by the Fresnel–Kirchhoff integral [see Eq.~\eqref{eq:2}].
However, in the backpropagation case, the propagation distance is taken with a minus sign, and the “reverse transmission” of the error field through a phase DOE is described by the multiplication of its complex amplitude by the complex conjugate of the CTF of this DOE~\cite{18, 19}.
Thus, the field $F_{m,j}( \mat{u}_{n + 1} ) = {\Pr_{{f_{n + 1}} \to f_m^+}}({F_{n + 1,j}})$ can be calculated recursively using the following formula:
\begin{equation}
\label{eq:14}
F_{l - 1,j}( \mat{u}_{l - 1} ) 
= C_l^*\iint {F_{l,j}( \mat{u}_l )T_l^*(\mat{u}_l)}\exp \left\{ { - \ii\frac{k_0}{2 d_l}{( \mat{u}_{l - 1} - \mat{u}_l )^2}} \right\}{\dd^2}{\mat{u}_l},\,\,\,l = n + 1,n,\ldots,m + 1,
\end{equation}
where for the calculation of the field $F_{n,j}( \mat{u}_n )$ (i.\,e., at $l=n+1$), one has to use ${T_{n + 1}}(\mat{u}_{n + 1}) \equiv 1$.
Let us note that since ${\Pr_{f_{n + 1} \to f_m^+ }}({\Delta_m}{w_{n + 1,j}}) = {\Delta_m}({w_{m,j}}{T_m})$, where $w_{m,j}({\mat{u}_m})T_m(\mat{u}_m)$ is the complex amplitude of the field immediately after the plane of the $m$-th DOE upon the forward propagation, the increment~\eqref{eq:13} can be transformed to the following form:
\begin{equation}
\label{eq:15}
{\Delta_m}{\varepsilon_j}\left( {\varphi_1,\ldots,\varphi_n} \right) 
= 2\operatorname{Re} \left\langle {{\Delta_m}\left( {{w_{m,j}}{T_m}} \right),{F_{m,j}}} \right\rangle  
= 2\operatorname{Re} \iint {{w_{m,j}}(\mat{u}_m)\Delta T_m(\mat{u}_m)F_{m,j}^*({\mat{u}_m}){\dd^2}\mat{u}_m}.
\end{equation}
Since
\begin{equation}
\label{eq:16}
\Delta {T_m} 
= \exp \left\{ \ii(\varphi_m + \Delta \varphi_m) \right\} - \exp \left\{ \ii \varphi_m \right\} 
= T_m \ii\Delta \varphi_m + {\rm o}\left( \Delta \varphi_m \right),
\end{equation}
the principal linear part of the increment~\eqref{eq:15} can be written as the following scalar product:
\begin{equation}
\label{eq:17}
{\delta_m}{\varepsilon_j}\left( \varphi_1, \ldots, \varphi_n \right) =  - 2\left\langle {\Delta {\varphi_m}({\mat{u}_m}),\operatorname{Im} \left[{w_{m,j}}({\mat{u}_m}){T_m}(\mat{u}_m)F_{m,j}^*(\mat{u}_m)\right]} \right\rangle .
\end{equation}
According to Eq.~\eqref{eq:17}, the Fréchet derivative of the functional of Eq.~\eqref{eq:6} has the form
\begin{equation}
\label{eq:18}
\frac{{\delta {\varepsilon_j}\left( {\varphi_1,\ldots,\varphi_n} \right)}}{{\delta {\varphi_m}}} =  -2\operatorname{Im} \left[w_{m,j}(\mat{u}_m) T_m({\mat{u}_m})F_{m,j}^*(\mat{u}_m)\right].
\end{equation}

Thus, the calculation of the gradient of the functional for a batch is carried out using Eqs.~\eqref{eq:8} and~\eqref{eq:18}.
Let us note that instead of the simple gradient descent method, one can use “improved” first-order methods such as the adaptive moment estimation method (ADAM)~\cite{32}.

\section{Design of robust DNNs}\label{sec:2}
In this section, we will consider a method for designing a DNN solving the classification problem assuming that the DOEs constituting the DNN are positioned in the optical system with some transverse errors.
Let us describe the positioning errors in the DOE planes by two-dimensional random vectors $\mat{U}_m^s = ( U_m^s, V_m^s),\,m = 1,\ldots,n$, where $U_m^s,V_m^s$ are random variables describing the transverse shifts of the $m$-th DOE along the coordinate axes $( u_m, v_m )$.
In this case, the complex transmission function of this DOE will have the form
\begin{equation}
\label{eq:19}
T_m^s({\mat{u}_m}) = T_m\left(\mat{u}_m - \mat{U}_m^s\right) = \exp \left\{ \ii \varphi_m \left(\mat{u}_m - \mat{U}_m^s\right) \right\},\,\,m = 1,\ldots,n.
\end{equation}
Note that for the sake of simplicity, we consider only the transverse shifts, since it is known that the shifts along the optical axis have a much smaller effect on the DOE performance~\cite{13}.

If the DOE CTFs have the form of Eq.~\eqref{eq:19}, the complex amplitudes of the fields $w_{m,j}( \mat{u}_m )$ generated upon forward propagation of the input field through the cascaded DOE will also depend on the positioning errors, i.\,e., will become functions of random vectors $\mat{U}_m^s = \left( U_m^s, V_m^s \right)$.
Indeed, these fields will be described by the recursive Eq.~\eqref{eq:2} with the DOE CTFs of Eq.~\eqref{eq:19}.
Accordingly, the error functionals of Eqs.~\eqref{eq:6} and~\eqref{eq:7} will also depend on random shifts $\mat{U}_m^s = ( U_m^s, V_m^s )$.
To emphasize this dependence, let us write these functionals as $\varepsilon ( \varphi_1, \ldots, \varphi_n; \mat{U}^s )$ and $\varepsilon_j( \varphi_1,\ldots,\varphi_n; \mat{U}^s )$, where $\mat{U}^s = ( \mat{U}_1^s,\ldots,\mat{U}_n^s )$ is a vector of dimension $2n$ composed of $n$ two-dimensional shift vectors $\mat{U}_m^s$.
At fixed phase functions $\varphi_1( \mat{u}_1 ),\ldots,\varphi_n( \mat{u}_n )$, the functionals turn into random variables.

Without loss of generality, let us assume that the error functionals take only nonnegative values.
Then, for the calculation of a DNN being robust to transverse shifts of the DOEs constituting it, it is necessary to minimize the expectation $\Phi ( \varphi_1, \ldots, \varphi_n ) = {\rm E}\left[ \varepsilon ( \varphi_1, \ldots, \varphi_n; \mat{U}^s ) \right]$.
Therefore, the problem of designing a robust DNN can be formulated as the problem of minimizing the functional
\begin{equation}
\label{eq:20}
\Phi ( \varphi_1, \ldots, \varphi_n ) = \sum\limits_{j = 1}^N {\rm E}\left[ \varepsilon_j\left( \varphi_1,\ldots,\varphi_n;\mat{U}^s \right) \right]  = \sum\limits_{j = 1}^N \Phi_j( \varphi_1,\ldots,\varphi_n ) 
\stackrel{\varphi_1, \ldots, \varphi_n}{\longrightarrow}
%\to \mathop 
\min .
%\limits_{\varphi_1, \ldots, \varphi_n} .
\end{equation}
Let the random vector $\mat{U}^s = ( \mat{U}_1^s,\ldots,\mat{U}_n^s)$ be defined by its probability density function $f( \mat{u}_1^s,\ldots,\mat{u}_n^s )$.
In this case, the expectations $\Phi_j( \varphi_1, \ldots, \varphi_n ) = {\rm E}\left[ \varepsilon_j\left( \varphi_1, \ldots, \varphi_n; \mat{U}^s \right) \right]$ take the form
\begin{equation}
\label{eq:21}
\Phi_j( \varphi_1,\ldots,\varphi_n ) 
= \iint \cdots\int \varepsilon_j\left( \varphi_1,\ldots,\varphi_n;\mat{u}_1^s,\ldots,\mat{u}_n^s \right)f\left( \mat{u}_1^s,\ldots,\mat{u}_n^s \right){\dd^2}\mat{u}_1^s\cdots{\dd^2}\mat{u}_n^s.
\end{equation}
The derivatives of the error functional~\eqref{eq:20} are equal to the sum of derivatives of the terms constituting it:
\begin{equation}
\label{eq:22}
\frac{{\delta \Phi ( \varphi_1,\ldots,\varphi_n)}}{{\delta {\varphi_m}}} = \sum\limits_{j = 1}^N {\frac{{\delta \Phi_j( \varphi_1,\ldots,\varphi_n )}}{{\delta \varphi_m}}}.
\end{equation}
Differentiating Eq.~\eqref{eq:21}, let us write the derivatives of the functional $\Phi_j(\varphi_1, \ldots, \varphi_n)$ as
\begin{equation}
\label{eq:23}
\frac{\delta \Phi_j( \varphi_1, \ldots, \varphi_n )}{\delta {\varphi_m}} = \iint\cdots\int \frac{\delta {\varepsilon_j}\left( {\varphi_1,\ldots,\varphi_n;\mat{u}_1^s, \ldots, \mat{u}_n^s} \right)}{\delta \varphi_m} f\left( {\mat{u}_1^s, \ldots, \mat{u}_n^s} \right){\dd^2}\mat{u}_1^s\cdots{\dd^2}\mat{u}_n^s.
\end{equation}
Note that the derivatives $\delta \varepsilon_j / \delta \varphi_m$ in Eq.~\eqref{eq:23} can be calculated using the previously obtained general Eq.~\eqref{eq:18} with the difference that the calculation of the fields $w_{m,j}(\mat{u}_m)$, $F_{m,j}(\mat{u}_m)$ occurring in this formula has to be carried out taking into account the DOE shifts, i.\,e.\ the CTFs of Eq.~\eqref{eq:19} have to be used.
According to Eq.~\eqref{eq:23}, the computation of the derivatives $\delta {\Phi_j} / \delta {\varphi_m}$ requires the calculation of a multiple integral in $2n$ dimensions, which is a computationally complex problem.
Therefore, for the computation of this integral, we propose to use the Monte Carlo method~\cite{33}.
In this method, $L$ independent realizations $\mat{u}_l^s = \left( \mat{u}_{1,l}^s, \ldots, \mat{u}_{n,l}^s \right),\,l = 1,\ldots,L$ of the random vector $\mat{U}^s = \left( \mat{U}_1^s,\ldots,\mat{U}_n^s \right)$ are considered.
Then, as an estimate of the derivative $\delta \Phi_j / \delta \varphi_m$, the following quantity is used:
\begin{equation}
\label{eq:24}
%{\rm Der}_{j,m}( \varphi_1, \ldots, \varphi_n ) =
\frac{\delta \Phi_j( \varphi_1, \ldots, \varphi_n )}{\delta {\varphi_m}} \approx
 \frac{1}{L}\sum_{l = 1}^L \frac{\delta {\varepsilon_j}\left( \varphi_1,\ldots,\varphi_n;\mat{u}_{1,l}^s,\ldots,\mat{u}_{n,l}^s \right)}{\delta \varphi_m}.
\end{equation}
According to Eq.~\eqref{eq:24}, for approximate calculation of the derivatives~\eqref{eq:23} of the functional~\eqref{eq:21}, it is necessary to calculate $L$ times the derivatives of the functional using Eq.~\eqref{eq:18} for randomly shifted DOEs with the CTFs $T_m^s(\mat{u}_m) = T_m\left(\mat{u}_m - \mat{u}_{m,l}^s\right),\,\,m = 1,\ldots,n$, and then average the obtained results.
Note that the presented estimate of the derivative~\eqref{eq:24} can be regarded as a realization of a random variable with the expectation coinciding with the exact value of the derivative $\delta \Phi_j( \varphi_1,\ldots,\varphi_n ) / \delta \varphi_m$ and the variance inversely proportional to the number of realizations~$L$.

Thus, in the present work, for the calculation of DNNs robust to transverse shifts, we propose to use a stochastic gradient descent method considered in the previous section, in which the calculation of the derivatives of the error functional is carried out using the approximate Eq.~\eqref{eq:24}.
	
It is worth noting that in addition to transverse shifts, other positioning errors can also be taken into account in the proposed method, for example, longitudinal shifts or random errors describing the DOE rotation around the optical axis.
Such errors could also be introduced in the input and output planes for describing the positioning errors of the input images of the objects being classified and of the detectors (target regions $G_k$).
Let us also note that in the method described above, instead of the error functional in the form of a mathematical expectation $\Phi ( \varphi_1, \ldots, \varphi_n ) = {\rm E}\left[ \varepsilon \left( \varphi_1,\ldots,\varphi_n; \mat{U}^s \right) \right]$ (the first moment), one can use a more complex criterion, for example, involving the second moment:
\begin{equation}
\label{eq:25}
   \Phi_2( \varphi_1,\ldots,\varphi_n ) = {\rm E}\left[ \varepsilon ^2\left( \varphi_1,\ldots,\varphi_n;\mat{U}^s \right) \right]
  = \sum\limits_{j = 1}^N {\rm E}\left[ \varepsilon_j^2\left( \varphi_1,\ldots,\varphi_n; \mat{U}^s \right) \right]  = \sum_{j = 1}^N \Phi_{2,j}( \varphi_1,\ldots,\varphi_n) .
\end{equation}
The derivatives of this functional can be calculated in a similar way using the Monte Carlo method.

\section{DNN design examples}\label{sec:3}
In this section, we consider several examples of designing robust and non-robust DNNs for the optical classification of handwritten digits from the MNIST database and present a comparison of their performance.

\subsection{DNNs consisting of a single DOE}
First, let us consider the solution of the classification problem using a single-DOE DNN.
For the design, the following parameters were chosen.
The fields $w_{0,j}(\mat{u}_0)$ representing the amplitude images of the input digits were defined in the center of the input plane on a $56\times56$ square grid with the step of $d=4\um$ at the incident wavelength $\lambda = 632.8\nm$.
Note that in the MNIST database, the images of the digits have the size of $28\times28$ pixels.
In order to define these images on a $56\times 56$ grid, they were interpolated according to the nearest value.
The phase function of the DOE was defined on a $512\times 512$ square grid with the same step of $d=4\um$.
In this case, the size (side length) of the square aperture of the DOE amounts to $2.048\mm$.
The distances between the input plane and the DOE and between the DOE and the output plane were taken to be the same and equal to $10\mm$.

\begin{figure}[hbt]
	\centering
		\includegraphics{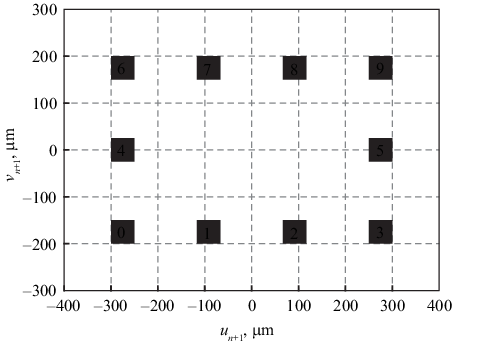}
	\caption{\label{fig:2} Geometry of the target regions in the output plane. }
\end{figure}

For solving the classification problem, 10~target regions $G_k$ were defined in the output plane.
These regions are shown in Fig.~\ref{fig:2} and are squares with the side length of $50\um$ placed along the boundary of a rectangular region with the size of $600\times 400 \um^2$.

As the error functionals $\varepsilon_j$ representing the errors of classifying different digits, we will use the cross entropy loss with the softmax function~\cite{7, 11, 16, 18}.
Under the energies $E_k$, we will understand the maximum values of the field intensity in the target regions [see Eq.~\eqref{eq:4}].
In this case, the functionals $\varepsilon_j$ will have the form
\begin{equation}
\label{eq:26}
\varepsilon_j( \varphi_1, \ldots, \varphi_n ) = D_j[E_1, \ldots, E_N] = -\ln \left[
\frac{\exp \left\{I_{n + 1,j}(\mat{u}_{n + 1,j})\right\}}
{\sum\limits_{k = 1}^N \exp \left\{I_{n + 1,j}(\mat{u}_{n + 1,k})\right\} } \right],
\end{equation}
where $E_k = I_{n + 1,j}(\mat{u}_{n + 1,k})$ are the intensity maxima in the target regions $G_k$ and $\mat{u}_{n + 1,k}$ are their coordinates.
The derivatives of the functional~\eqref{eq:26} are calculated using the general formula~\eqref{eq:18}, where the function $F_{m,j}( \mat{u}_m )$ is calculated through the backpropagation of the error field defined by Eq.~\eqref{eq:12}.
By substituting Eq.~\eqref{eq:26} into Eq.~\eqref{eq:12}, we obtain the error field as
\begin{equation}
\label{eq:27}
%\begin{aligned}
  F_{n + 1,j}( \mat{u}_{n + 1}) = \frac{w_{n + 1,j}(\mat{u}_{n + 1})}{2}\Bigg[ - \delta (\mat{u}_{n + 1} - \mat{u}_{n + 1,j})    
%		\\&
		+ \frac{\exp \left\{I_{n + 1,j}(\mat{u}_{n + 1})\right\}}{S(\mat{u}_{n + 1})}  \sum_{k = 1}^N {\delta (\mat{u}_{n + 1} - \mat{u}_{n + 1,k})} \Bigg], 
%\end{aligned}
\end{equation}
where 
$$
S(\mat{u}_{n + 1}) = \sum_{k = 1}^N \exp \left\{I_{n + 1,j}(\mat{u}_{n + 1,k})\right\}.
$$

\begin{figure}[hbt]
	\centering
		\includegraphics{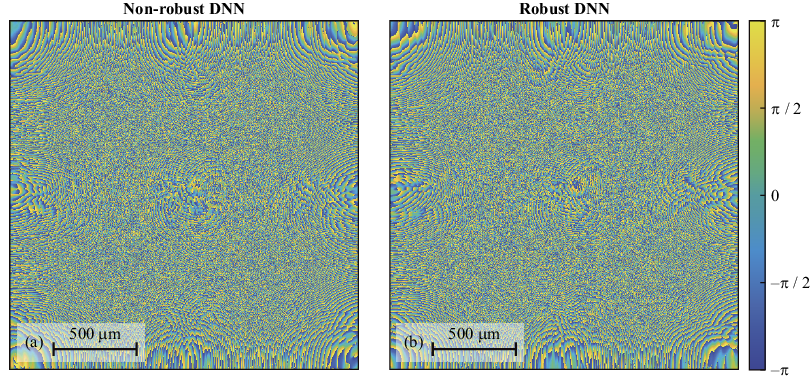}
	\caption{\label{fig:3} Phase functions of non-robust (a) and robust (b) single-DOE DNNs. }
\end{figure}

In order to assess the influence of the transverse shifts on the DOE operation, we first designed a non-robust DOE using the stochastic gradient descent method without taking into account these shifts.
For the calculation, a training set containing 60000 images of the digits was used.
The training was carried out using batches, each containing 20~randomly chosen images of digits.
The derivatives of the error functional~\eqref{eq:7},~\eqref{eq:26} were calculated using Eqs.~\eqref{eq:18} and~\eqref{eq:27}.
Let us note that the calculation of the fields used in the expressions for the derivatives was performed using the angular spectrum method~\cite{34, 35}.
In the DOE calculation, 12000 iterations were carried out, which took of about 100~minutes on a desktop PC (Intel Core i9-10920X CPU @ 3.50GHz).
The obtained phase function of the DOE is shown in Fig.~\ref{fig:3}(a).

\begin{figure}[hbt]
%	\centering
	\hspace{-4em}
		\includegraphics[scale=0.85]{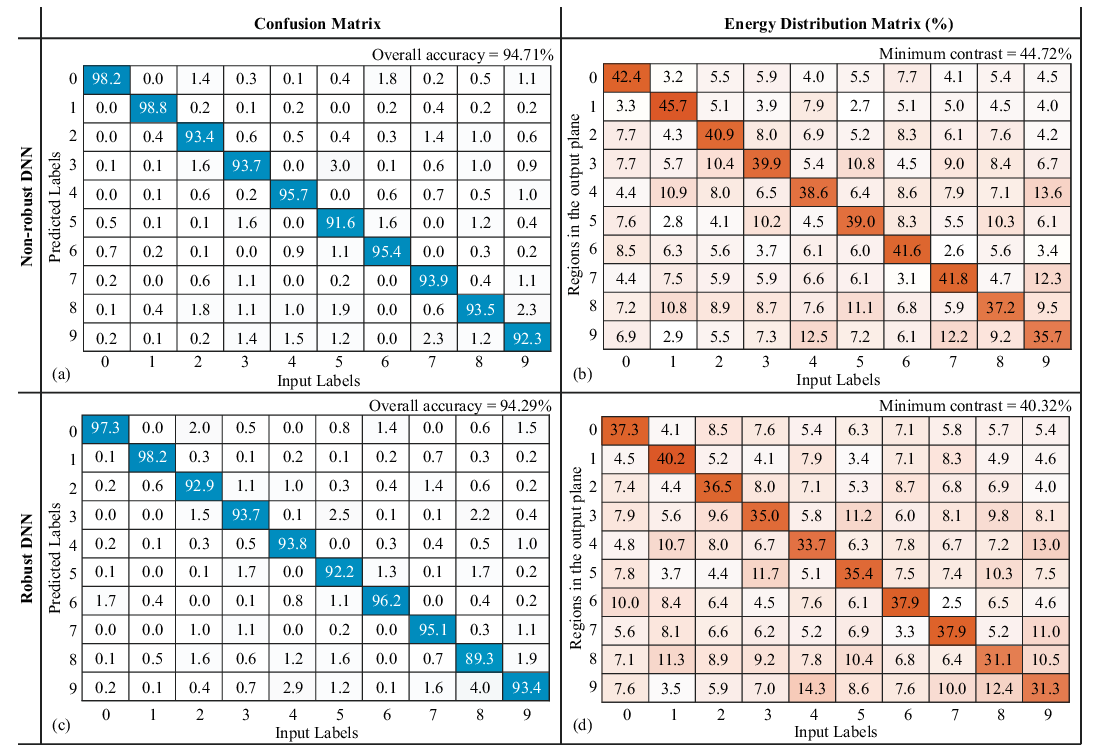}
	\caption{\label{fig:4} Confusion matrices~(a),~(c) and energy distribution matrices~(b),~(d) for non-robust~(a),~(b) and robust~(c),~(d) single-DOE DNNs. }
\end{figure}

After training, the DOE performance was evaluated on a test set containing 10000 images of digits not included in the training set.
The obtained results presented as the confusion matrix and the energy distribution matrix are shown in Figs.~\ref{fig:4}(a) and~\ref{fig:4}(b).
The element $(j, k)$ of the confusion matrix shows the percentage of the objects of the $j$-th class (i.\,e., images of the digit “$j$”), which were classified as the elements of the $k$-th class (i.\,e., as images of the digit “$k$”).
Thus, the diagonal elements of this matrix represent the classification accuracy values for different digits.
Let us note that in the theory of neural networks, this quantity is usually referred to as recall.
Over the confusion matrix, the overall classification accuracy (i.\,e., the ratio of the correctly classified digits to the size of the test set) is presented, which equals 94.71\% in the considered case.

The element $(j, k)$ of the energy distribution matrix represents average energy calculated over the test set, which is generated in the $k$-th target region for input objects of the $j$-th class.
From the practical point of view, it is important to assess the energy in the region corresponding to the class under consideration in comparison with the energies in the regions corresponding to the other classes.
To do this, we will use the following quantity, which can be referred to as contrast for the $j$-th class:
\begin{equation}
\label{eq:28}
K_j = \frac{{\bar E}_{j,j} - \mathop {\max }\limits_{k \ne j} {{\bar E}_{j,k}}}{{\bar E}_{j,j} + \mathop {\max }\limits_{k \ne j} {{\bar E}_{j,k}}} \times 100\%,
\end{equation}
where $\bar E_{j,k}$ are the elements of the energy distribution matrix.
In the opinion of the authors, for stable recognition of “true maxima” of the energies in the experimental implementation of the DNN, it is necessary for the theoretically obtained $K_j$ values to exceed at least 10\%.
Over the energy distribution matrix [Fig.~\ref{fig:4}(b)], the minimum contrast value $K_{\rm min} = \min_j K_j$ is presented, which in the considered case amounts to 45.72\% and significantly exceeds the chosen threshold value.

\begin{figure}[bth]
	\centering
		\includegraphics{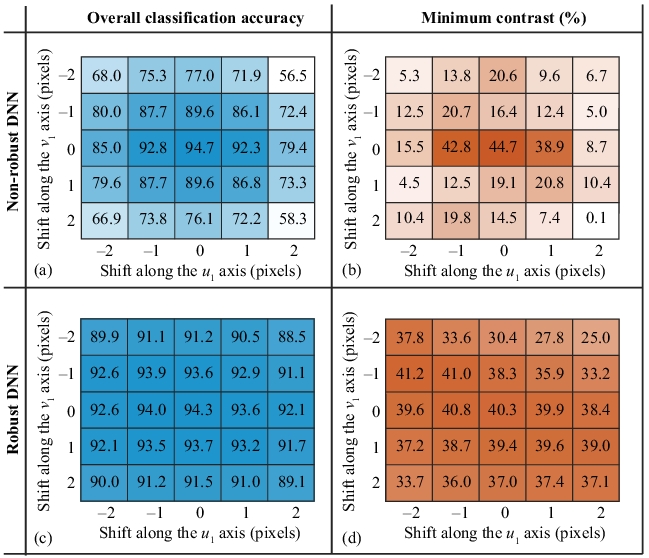}
	\caption{\label{fig:5}Overall classification accuracy~(a),~(c) and minimum contrast~(b),~(d) at different transverse shifts for non-robust~(a),~(b) and robust~(c),~(d) single-DOE DNNs. }
\end{figure}

Let us now investigate, how the single-DOE DNN performance changes when transverse shifts are introduced.
Figures~\ref{fig:5}(a) and~\ref{fig:5}(b) show the values of the overall classification accuracy and minimum contrast calculated at different DOE shifts defined by the following vectors:
\begin{equation}
\label{eq:29}
\mat{u}_{1,i,j}^s = (id,jd),\,\,\,i,j = -2, -1,0,1,2,
\end{equation}
where $d=4\um$ is the pixel size (discretization step of the phase function of the DOE).
The corresponding indices $(i, j)$ defining the magnitudes of the shifts along the coordinate axes are presented in the captions to the rows and columns of the tables in Figs.~\ref{fig:5}(a) and~\ref{fig:5}(b).
Note that the central elements with the indices $i=j=0$ correspond to the overall accuracy and minimum contrast for a non-shifted DOE.
For the calculation of the elements of the tables at $i \ne 0,\,j \ne 0$, the designed DOE [Fig.~\ref{fig:3}(a)] was shifted by a vector of Eq.~\eqref{eq:29} (i.\,e., its CTF was assumed to be equal to $T_1^s(\mat{u}_1) = \exp \left\{ \ii\varphi_1\!\left(\mat{u}_1 - \mat{u}_{1,i,j}^s\right) \right\}$) and then the overall classification accuracy and minimum contrast for the test set were calculated.
Let us call the shifts by one and two pixels the sets of shift vectors of Eq.~\eqref{eq:29}, for which $\max \left\{|i|, |j|\right\} = 1$ and $\max \left\{|i|, |j|\right\} = 2$, respectively.
One can see that in the case of a shift by 1~pixel (eight cells of the tables around the central cell $i = j = 0$), the overall accuracy decreases to 86.1\% and the minimum contrast becomes 12.4\%.
In the case of a shift by 2~pixels (16~outer cells of the tables), the overall accuracy decreases further (the minimum value drops to 56.5\%), whereas the minimum contrast in half of the cells becomes less than the threshold value of 10\%.
Therefore, one can conclude that a shift of the DOE by two pixels makes it inoperable.

Then, using the stochastic gradient descent method, a robust single-DOE DNN was calculated taking into account the transverse shifts.
In the calculation, the shifts were described by a continuous random vector uniformly distributed in the square $[-d, d]\times [-d,d]$. 
In accordance with the description of the method presented in Section~\ref{sec:3}, for calculating the derivatives of the error functional~\eqref{eq:22},~\eqref{eq:23}, for each batch, $L=4$ independent realizations $\mat{u}_{1,l}^s,\,l = 1,2,3,4$ of the discrete random vector $\mat{U}_1^s = ( U_1^s, V_1^s )$ were generated.
For these realizations, for DOEs with the CTFs $T_1^s(\mat{u}_1) = \exp \left\{ {\ii\varphi_1\!\left(\mat{u}_1 - \mat{u}_{1,l}^s\right)} \right\},\,l = 1,2,3,4$, derivatives of the error functionals~\eqref{eq:26} were calculated using Eqs.~\eqref{eq:18} and~\eqref{eq:27} and then the estimates of the derivatives of the error functionals~\eqref{eq:23} and~\eqref{eq:22} were found using Eq.~\eqref{eq:24}. 
The phase function of the DOE calculated using the stochastic gradient method with the derivatives calculated in this way is shown in Fig.~\ref{fig:3}(b). 
Figures~\ref{fig:4}(c) and~\ref{fig:4}(d) show the confusion matrix and energy distribution matrix for this DOE for the used test set. 
One can see that the overall classification accuracy and the minimum contrast for the designed DOE amount to 94.29\% and 40.32\%, respectively. 
In comparison with the non-robust DOE, the obtained values are slightly smaller, but their decrease is rather insignificant and amounts to 0.42\% and 4.4\% for the overall accuracy and minimum contrast, respectively. 
Figures~\ref{fig:5}(c) and~\ref{fig:5}(d) show the overall accuracy and minimum contrast calculated at different transverse shifts of the DOE defined by Eq.~\eqref{eq:29}.
As before, for the calculation of these tables, the designed DOE [Fig.~\ref{fig:3}(b)] was shifted by a vector of Eq.~\eqref{eq:29} and then the values of the overall classification accuracy and minimum contrast were calculated for the test set.
By comparing Fig.~\ref{fig:5}(a) with Fig.~\ref{fig:5}(c) and Fig.~\ref{fig:5}(b) with Fig.~\ref{fig:5}(d), one can see that, as expected, the transverse shifts have a significantly less effect on the performance of the robust DOE.
In particular, at two-pixel shifts, the minimum overall accuracy and contrast amount to 88.5\% and 25\%, respectively (as compared to 56.5\% and 0.1\% for the non-robust DOE).

In what follows, for comparing the performance of robust and non-robust DNNs, we will use the values of the average classification accuracy and average minimum contrast calculated at DOE shifts by one and two pixels.
These average values were calculated using the data shown in Fig.~\ref{fig:5} and are presented in Table~\ref{tab:1}.
Note that the “No shifts” column contains the central elements of the tables presented in Fig.~\ref{fig:5}.
According to the definitions of the shifts by~1 and 2~pixels given above, the column “Shift by 1~pixel” contains average values of the overall classification accuracy and minimum contrast calculated over eight cells of the table around the central cell with the indices $i = j = 0$, whereas the column “Shift by 2~pixels” contains the corresponding values averaged over the rest (outer) cells of the tables.
The results in Table~\ref{tab:1} demonstrate that while the robust DOE is very slightly inferior in its performance if the shifts are not present, it ensures significantly higher average accuracy and average minimum contrast in the case of shifts by~1~and 2~pixels.
In particular, in the case of shifts by 2~pixels, the robust DOE provides a 18.1\% greater average classification accuracy and a more than 3~times greater average minimum contrast.

\begin{table}[]
\caption{\label{tab:1}Classification accuracy and minimum contrast for non-robust and robust single-DOE DNNs at different shifts.}
\centering
\begin{tabular}{cccc}
\hline\hline
\multirow{2}{*}{Type of the designed DNN} & \multicolumn{3}{c}{Average accuracy and average minimum contrast (\%)} \\ \cline{2-4} & 
\parbox{2.5cm}{\centering No shifts} & 
\parbox{2.5cm}{\centering Shift by 1 pixel} & 
\parbox{2.5cm}{\centering Shift by 2 pixels} \\ \hline
Non-robust & (94.7; 44.7) & (89.1; 23.0) & (72.9; 10.3) \\
Robust & (94.3; 40.3)  & (93.6; 39.2) & (91.0; 33.9) \\ \hline\hline
\end{tabular}
\end{table}

\subsection{DNNs consisting of a cascade of two DOEs}
\begin{figure}%[hbt]
	\centering
		\includegraphics{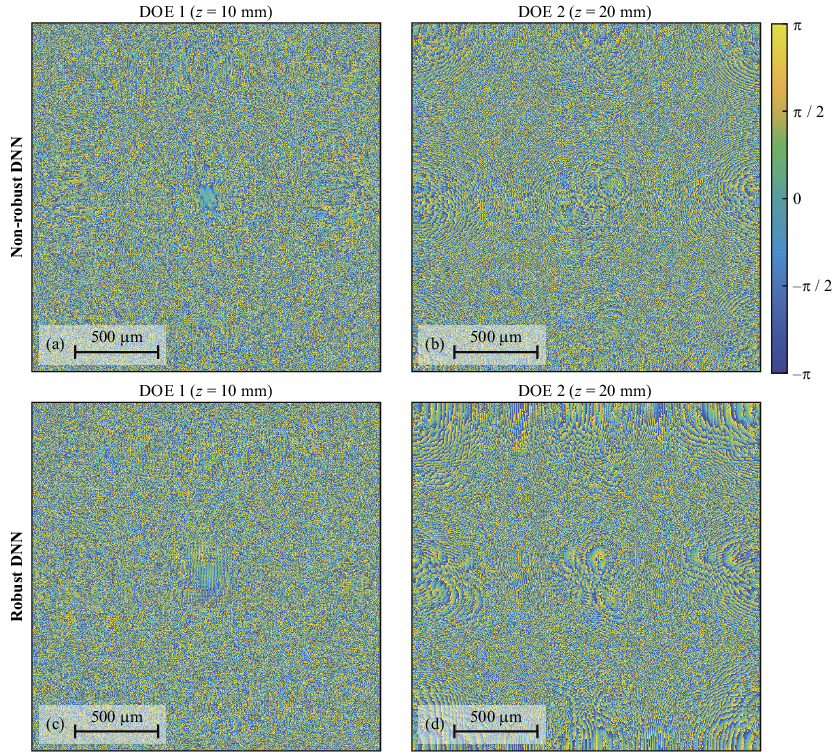}
	\caption{\label{fig:6} Phase functions of non-robust~(a),~(b) and robust~(c),~(d) two-DOE DNNs.}
\end{figure}

\begin{figure}%[hbt]
	%	\centering
	\hspace{-4em}
		\includegraphics[scale=0.85]{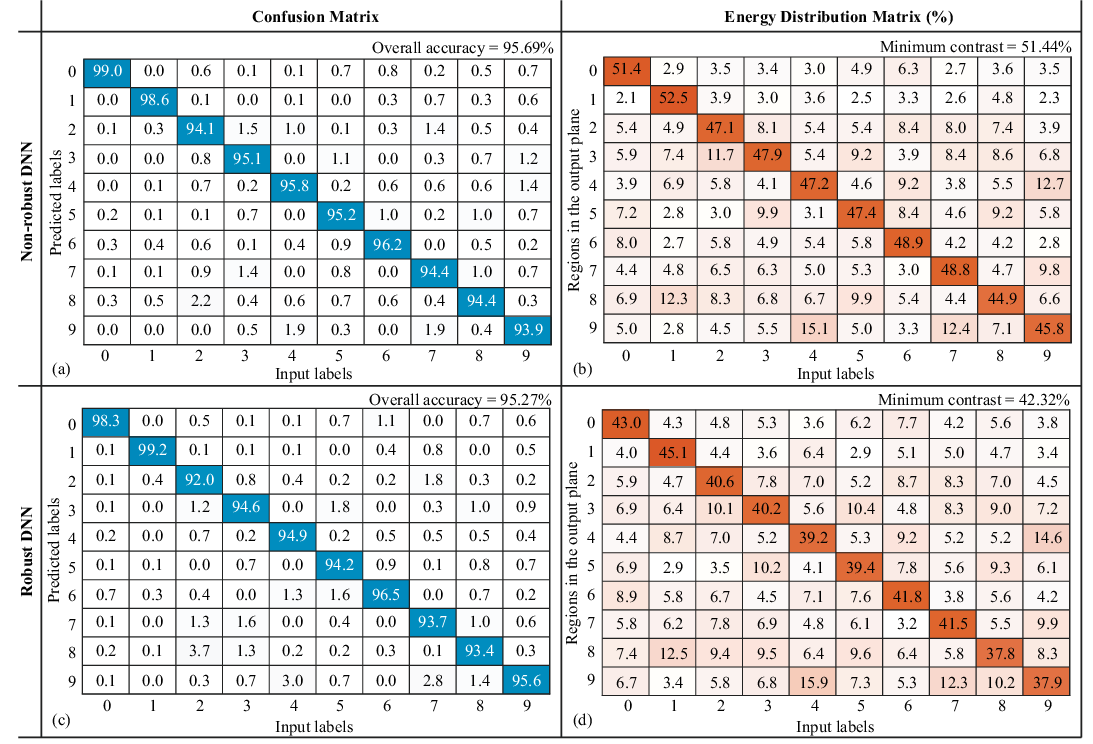}
	\caption{\label{fig:7} Confusion matrices~(a),~(c) and energy distribution matrices~(b),~(d) for non-robust~(a),~(b) and robust~(c),~(d) two-DOE DNNs.}
\end{figure}

As the second example, we designed DNNs consisting of a cascade of two DOEs located in the planes $z = f_1 = 10\mm$ and $z = f_2 = 20\mm$.
In this case, the output plane is located at $z = 30\mm$.
All other parameters (discretization step, operating wavelength, aperture size) coincide with the parameters of the examples considered above.
The phase functions of the calculated cascaded DOEs are shown in Fig.~\ref{fig:6}.
As in the previous section, non-robust and robust DNNs were calculated.
When calculating the robust DNN, the shift vectors of both DOEs constituting the DNN were described by random vectors uniformly distributed in a square with the side of $2d$.
The results of testing the designed DNNs (without the positioning errors) in the form of confusion matrices and energy distribution matrices are shown in Fig.~\ref{fig:7}.
As before, for the testing, a set containing 10000 images not included in the training set was used.
One can see that moving to the two-DOE configuration enables, as compared to a single DOE, to increase the overall classification accuracy by almost 1\% while simultaneously increasing the minimum contrast.
Without the transverse shifts present, the non-robust DNN has a slightly better (by 0.4\%) overall classification accuracy and minimum contrast better by more than 9\%.

Let us now study, how the performance of the designed DNNs changes when transverse shifts of the DOEs are introduced.
Upper part of Table~\ref{tab:2} shows the average values of the classification accuracy and minimum contrast for the non-robust DNN at different transverse shifts of the DOEs.
In the cell, which corresponds to the case when both DOEs are not shifted, the values of the overall classification accuracy and minimum contrast from Figs.~\ref{fig:7}(a) and~\ref{fig:7}(b) are given.
Let us explain the values in the other cells of the table, considering the cell describing the case, in which both the first and the second DOE constituting the DNN are shifted by 1 pixel.
Let us remind that under the shift by one pixel, we understand a set of eight shift vectors of Eq.~\eqref{eq:29}, for which $\max \left\{|i|, |j|\right\} = 1$.
Therefore, for calculating the overall accuracy and minimum contrast for the case of shifting both DOEs by 1~pixel, a total of 64~different arrangements of the 1-st and 2-nd DOEs were considered, each DOE having its “own” shift by one pixel. 
For each such arrangement, the overall classification accuracy and minimum contrast for the used test set were calculated, and then averaged over the 64 considered cases. 
The average values obtained in this way are presented in the corresponding cell of the table.
The data presented in the rest of the table cells were calculated in a similar way.

From Table~\ref{tab:2}, one can see that a non-robust DNN consisting of two DOEs is very sensitive to these shifts.
In particular, if the first DOE is shifted by 1~pixel, and the second DOE is not shifted, the average classification accuracy decreases already to 79.12\% at average minimum contrast of 18.16\%. 
If both DOEs are shifted by 1~pixel, the DNN becomes almost inoperable: the average classification accuracy drops to 66.11\%, whereas the average minimum contrast becomes less than 10\%.
Let us remind that in the case of a non-robust single-DOE DNN, the performance decrease caused by a one-pixel shift was noticeably smaller (see Table~\ref{tab:1}).

Average values of the classification accuracy and minimum contrast for the designed robust two-DOE DNN are given in the lower part of Table~\ref{tab:2}.
By comparing the upper and lower parts of Table~\ref{tab:2}, one can see that the robust DNN, in contrast to the “usual” (non-robust) DNN, exhibits good performance when both DOEs are shifted by~1 or even 2~pixels.
It is worth emphasizing that when both DOEs are shifted by 2~pixels, the average classification accuracy remains greater than 93\%, and the minimum average contrast is almost 40\%.
Therefore, the presented DNN design approach taking into account the positioning errors of the DOEs constituting the DNN enables compensating for %errors corresponding to
transverse shifts by 1--2~pixels (4--$8\um$) along both coordinate axes.
These values correspond to the shift of the DOE centers by distances of up to %from~6 to~
17 wavelengths of the incident radiation.

\begin{table}
	\caption{\label{tab:2}Average classification accuracy and average minimum contrast for non-robust and robust two-DOE DNNs.}
	\centering
	\begin{tabular}{cccc}
	\hline\hline
	\multirow{2}{*}{Shift of the 2-nd DOE} & \multicolumn{3}{c}{Shift of the 1-st DOE} \\ \cline{2-4} & 
	\parbox{2.5cm}{\centering No shift} & 
	\parbox{2.5cm}{\centering 1 pixel} & 
	\parbox{2.5cm}{\centering 2 pixels} \\ \hline
	\multicolumn{4}{c}{\textbf{Non-robust DNN}}\\
	No shift & (95.69; 51.45) & (79.12; 18.16) & (44.41; 4.81)\\
	1 pixel  & (72.08; 10.02) & (66.11; 8.30) & (44.17; 5.51)\\
	2 pixels & (52.23; 5.09) &	(47.56; 4.82) & (36.35; 3.89)\\ \hline

\multicolumn{4}{c}{\textbf{Robust DNN}}\\
	No shift & (95.27; 42.33) & (95.13; 42.62) & (94.20; 41.80) \\
	1 pixel  & (95.21; 42.59) & (94.98; 42.03) & (94.08; 41.16) \\
	2 pixels & (94.73; 42.06) & (94.40; 41.33) & (93.34; 39.70) \\\hline\hline
	\end{tabular}
\end{table}

To conclude this section, let us note that we also investigated DNNs consisting of a cascade of three DOEs (not presented in the paper for the sake of brevity).
We found that adding a third DOE virtually does not increase the classification accuracy and only slightly (by 1--2\%) increases the contrast values.
The investigation has shown that a non-robust three-DOE DNN is even more sensitive to DOE shifts and becomes inoperable even when one of the three DOEs is shifted by 1~pixel.
At the same time, the robust DNN still retains good performance when all three DOEs are shifted by 2~pixels.

\section{Conclusion}
In this work, we considered a gradient method for the design of robust diffractive neural networks consisting of a cascade of phase DOEs for image classification, which takes into account the positioning errors (random transverse shifts) of the DOEs.
In the proposed method, the error of solving the classification problem was first represented by a functional depending not only on the phase functions of the cascaded DOE, but also on random vectors describing the transverse shifts of the DOEs.
Then, the mathematical expectation of this functional, depending only on the phase functions of the DOEs, was used as an error functional in the problem of calculating the robust DNN taking into account the transverse shifts.
Explicit expressions were obtained for the derivatives of this error functional.
It was shown that the calculation of the derivatives of this functional using the Monte Carlo method corresponds to the DNN training method, in which the DOEs are positioned with random transverse shifts.

Using the proposed gradient method, we designed robust and non-robust DNNs operating at a wavelength of $632.8\nm$ and consisting of one and two DOEs with a pixel size (DOE discretization step) of $4\um$.
The numerical simulation results demonstrated that the robust DNNs retain good performance (classification accuracy exceeding 93\% and minimum contrast of almost 40\%) with two-pixel shifts along both coordinate axes, which corresponds to a shift of the DOE center by a distance of up to 17~wavelengths.
At the same time, the non-robust two-DOE DNN turns out to be virtually inoperable even when the DOEs are shifted by a single pixel.

It is important to note that, along with transverse shifts, other positioning errors can also be taken into account in the proposed method, in particular, longitudinal shifts or DOE rotations around the optical axis.
In addition, the proposed design method can be used to calculate robust DNNs and cascaded DOEs intended for solving other problems, in particular, for the optical implementation of mathematical transformations described by linear operators.
In this case, the modification of the method will consist only in changing the used error functional: instead of a functional describing the classification error, one should use a functional describing the error of calculating the required linear operator.

\section{Acknowledgments}
This work was carried out within the state assignment by the Ministry of Science and Higher Education of the Russian Federation to Samara University (project FSSS-2024-0014; development
of a gradient method for calculating robust DNNs, design and investigation of DNN examples) and was supported by the Russian Science Foundation (grant 24-19-00080; general methodology for calculating the Fr\'echet derivatives of the error functionals based on the unitarity property of light propagation operators).

\bibliographystyle{elsarticle-num} 
\bibliography{Robust}

\end{document}